%% file: kev_stringaxion_v1.4.tex
\documentclass[aps,prd,reprint,superscriptaddress,nofootinbib,longbibliography,preprintnumbers,amsfonts,amsmath,amssymb]{revtex4-1}

\usepackage{ifpdf}
\usepackage{braket}

\usepackage{ulem} %allows strikethrough text with \sout{...}
\usepackage{slashed}
\usepackage{cancel}

\ifpdf
    \usepackage[pdftex,colorlinks=true,plainpages=false,pdfpagelabels]{hyperref}
    \usepackage[pdftex]{color}
    \hypersetup{colorlinks = true}
\else
    \usepackage[colorlinks=true]{hyperref}
\fi

\begin{document}

\preprint{IPMU14-0240, UCB-PTH-14/33}

\title{A keV String Axion from High Scale Supersymmetry}

\author{Brian Henning}
  \email{bhenning@berkeley.edu}
  \affiliation{Department of Physics, University of California, Berkeley,
               California 94720, USA}
  \affiliation{Theoretical Physics Group, Lawrence Berkeley National Laboratory, Berkeley,
               California 94720, USA}

\author{John Kehayias}
  \email{john.kehayias@ipmu.jp}
  \affiliation{Kavli Institute for the Physics and Mathematics of the Universe (WPI)\\
               Todai Institutes for Advanced Study, The University of Tokyo\\
               Kashiwa, Chiba 277-8582, Japan}

\author{Hitoshi Murayama}
  \email{hitoshi.murayama@ipmu.jp}
  \affiliation{Department of Physics, University of California, Berkeley,
               California 94720, USA}
  \affiliation{Theoretical Physics Group, Lawrence Berkeley National Laboratory, Berkeley,
               California 94720, USA}
  \affiliation{Kavli Institute for the Physics and Mathematics of the Universe (WPI)\\
               Todai Institutes for Advanced Study, The University of Tokyo\\
               Kashiwa, Chiba 277-8582, Japan}

\author{David Pinner}
  \email{dpinner@berkeley.edu}
  \affiliation{Department of Physics, University of California, Berkeley,
               California 94720, USA}
  \affiliation{Theoretical Physics Group, Lawrence Berkeley National Laboratory, Berkeley,
               California 94720, USA}

\author{Tsutomu T.~Yanagida}
  \email{tsutomu.tyanagida@ipmu.jp}
  \affiliation{Kavli Institute for the Physics and Mathematics of the Universe (WPI)\\
               Todai Institutes for Advanced Study, The University of Tokyo\\
               Kashiwa, Chiba 277-8582, Japan}

\begin{abstract}
%%%%%%%%%%%%%%%%%%%%%%%%%%%%%%%%%%%%%%%%%%%%%%%%%%%%%%%%%%%%%%%%%%%%%%%%%%%%%%%%%

  \noindent
  Various theoretical and experimental considerations motivate models
  with high scale supersymmetry breaking. While such models may be
  difficult to test in colliders, we propose looking for signatures at
  much lower energies. We show that a keV line in the X-ray spectrum
  of galaxy clusters (such as the recently disputed 3.5 keV
  observation) can have its origin in a universal string axion coupled
  to a hidden supersymmetry breaking sector. A linear combination of
  the string axion and an additional axion in the hidden sector
  remains light, obtaining a mass of order 10 keV through
  supersymmetry breaking dynamics. In order to explain the X-ray line,
  the scale of supersymmetry breaking must be about $10^{11-12}$
  GeV. This motivates high scale supersymmetry as in pure gravity
  mediation or minimal split supersymmetry and is consistent with all
  current limits. Since the axion mass is controlled by a dynamical
  mass scale, this mass can be much higher during inflation, avoiding
  isocurvature (and domain wall) problems associated with high scale
  inflation. In an appendix we present a mechanism for dilaton
  stabilization that additionally leads to $\mathcal{O}(1)$
  modifications of the gaugino mass from anomaly mediation.

%%%%%%%%%%%%%%%%%%%%%%%%%%%%%%%%%%%%%%%%%%%%%%%%%%%%%%%%%%%%%%%%%%%%%%%%%%%%%%%%%
\end{abstract}

\maketitle

\section{Introduction}

The hierarchy problem has dominated much of the discussion on physics
beyond the Standard Model (SM) in the past three decades, and
supersymmetry emerged as the leading contender to solve this
problem. In order to solve the problem fully, there was much
anticipation that supersymmetry should be discovered very soon after
the LHC began operating. Unfortunately, the LHC Run-I at 7--8~TeV
placed a very strong lower limit, typically above a TeV, on
superparticle masses~\cite{atlas_susy, *cms_susy}, even though the
quantitative limits are quite sensitive to the assumptions on the mass
spectrum as well as the decay modes.

In addition, the discovered mass of the Higgs boson at
125~GeV~\cite{atlashiggs,*cmshiggs} is higher than what was expected
in the Minimal Supersymmetric Standard Model (MSSM). If we rely on the
radiative corrections~\cite{higgs1,*higgs2,*higgs3,*higgs4} from
superparticles to raise the mass of the Higgs boson, we need to place
scalar top quarks above a TeV. Finally, there have been long standing
issues with supersymmetry, such as the absence of effects from large
flavor-changing neutral currents, cosmological problems with the
gravitino, and string moduli, which all prefer a supersymmetry
spectrum with scalars around $m_{\text{SUSY}}
\approx$~100--1000~TeV. If we take these hints seriously, direct
searches for supersymmetry at collider experiments will be very
difficult in the foreseeable future.

It is important to ask the question of whether there are alternative
ways to find an experimental hint for supersymmetry. We argue in this
letter that the energy scale $m_\mathrm{SUSY}^2/M_{\text{pl}}
\approx$~keV may provide us with an indirect window to supersymmetry
beyond the reach of accelerator experiments. Here \(M_{\text{pl}}\) is
the reduced Plank scale, \(M_{\text{pl}} \simeq 2.4 \times 10^{18}\) GeV.

The recent observation of an unidentified line at about 3.5 keV in the
X-ray spectrum of galaxy clusters~\cite{exp1,exp2} hints at new
particles at the keV energy scale. Although it has since been disputed
by several other (non-) observations~\cite{Malyshev:2014xqa,
  *Anderson:2014tza, *Tamura:2014mta}, it is interesting to consider
that it (or a line observed in the future) could be a signal of dark
matter decaying into photons. Even if we attribute this particular
line to astrophysical processes, looking for new lines in X-rays is a
continuing prospect. The possibility of linking such a low energy
signal to physics at very high scales is an intriguing new avenue that
we will explore in this work, using the 3.5 keV line as our guiding
example. However, the types of models we will consider are rather
generic and are not tied only to this specific experimental result.

Inspired by this observation, we investigate how supersymmetry may be
relevant to the observed excess in X-rays from clusters of
galaxies. Given the monochromatic line feature, it is tempting to
consider a dark matter particle decaying into two photons.  Note the
Landau--Yang theorem that a vector cannot decay into two photons. Thus
we consider a scalar particle decaying into two
photons.\footnote{However, there may be an alternative possibility
  that a fermion, such as a sterile neutrino, decays into a light
  active neutrino and a photon, through a suppressed mixing between
  the sterile and active neutrinos (see e.g.~the review
  \cite{neutrinokusenko}).} Then we need to understand the radiative
stability of the keV energy scale, in addition to the origin of the
keV scale itself.

%\paragraph{Orders of magnitude}
%--- 
The minute we assume that $m_\mathrm{SUSY}$ may be around 1000~TeV,
there is a possible derived energy scale of
$m_\mathrm{SUSY}^2/M_{\text{pl}} \approx$~keV. One immediate
possibility that comes to mind is that $m_\mathrm{SUSY}$ may be the
scale of supersymmetry breaking itself, such as in low-energy gauge
mediation~\cite{Hamaguchi:2014sea}, and keV is the mass scale of the
gravitino or moduli. This possibility was examined already in the
literature. For example, the gravitino~\cite{Moroi:1993mb} or
moduli~\cite{deGouvea:1997tn,moduli2,moduli3} may be dark matter. The
decay of the moduli in this context may produce an X-ray signal from
the clusters of galaxies~\cite{moduli1,Kusenko:2012ch}. However, there
are several non-trivial problems in gauge mediation, such as the
$\mu$-problem, the overproduction of gravitinos, and producing the
correct Higgs boson mass (there are consistent models evading such
difficulties, though, as in~\cite{Hamaguchi:2014sea}).

We point out in this letter that there is an alternative possibility.
$m_\mathrm{SUSY} \approx$ 100--1000~TeV may be the gravitino
mass. This possibility has attracted quite a bit of interest in the
literature recently, starting from anomaly
mediation~\cite{anomalymed1, *anomalymed2, *anomalymed3} and leading
up to pure gravity
mediation~\cite{puregravity1,*puregravity2,*puregravity3} or minimal
split supersymmetry~\cite{splitsusy}. In this case the scale of
supersymmetry breaking is $\Lambda_{\text{SUSY}} \approx
(m_{\text{SUSY}} M_{\text{pl}})^{1/2} \approx 10^{12}$~GeV. The keV
scale emerges parametrically as $\Lambda_\mathrm{SUSY}^4/M_{\text{pl}}^3$.

%\paragraph{Scalar properties}
%--- 
If the new particle is a scalar, the keV mass scale must be protected
against radiative corrections. The most effective mechanism is if the
particle is a pseudo-Nambu--Goldstone-boson (pNGB). We call it
generically an \textit{axion} even though it may not have anything to
do with the solution to the strong CP problem of QCD. The possibility
that a pNGB may explain the origin of the 3.5 keV photon line has been
also been considered elsewhere~\cite{kevaxion1, *kevaxion2,
  *kevaxion3, *kevaxion4}. In this letter, we point out how such a pNGB can have a
natural origin in the context of high scale supersymmetry.

The scalar decay proceeds through a dimension five operator suppressed
by scale \(M\)
with a rate \(\Gamma \approx m^3/8\pi M^2\).
For a 7~keV particle, the observed decay rate\footnote{Of course, by
  changing slightly the scale $M$ the rate can be below current
  experimental bounds.} is well-described by the energy scale
$M \approx 0.1M_{\text{pl}}$. If interpreted as the axion decay
constant $M = 32\pi^2 f$, $f \approx 10^{15}$~GeV. Therefore,
discussing only two important scales, $\Lambda_{SUSY}$ and
$M_{\text{pl}}$, seems well-warranted.

%What is the origin of this scalar particle?
Given the large scale \(M\)
and the coupling to electromagnetism, a well motivated possibility is
to consider the scalar to be a modulus or axion field from string
theory, where such properties can occur naturally,
e.g.~\cite{stringaxion}. We will consider the ``universal'' or
model-independent string axion, the defining properties of which we
take to be the high scale decay constant and a universal coupling to
all $F\widetilde{F}$. Thus the string axion couples to the hidden sector
responsible for dynamical SUSY breaking and it may have a mass
\(m \sim \Lambda_{\text{SUSY}}^2/f\).
However, \textit{this is not the case if the hidden sector contains an
  anomalous, global} \(U(1)\)
\textit{symmetry that is spontaneously broken}. In this case, a second
axion emerges which mixes with the string axion and leaves a massless
eigenstate. Note that a spontaneously broken, anomalous \(U(1)\)
is a common feature of dynamical SUSY breaking models; the necessity
of lifting flat directions in order to break supersymmetry typically
induces non-zero vacuum expectation values, thus breaking global
symmetries.

In the above scenario, where the SUSY breaking sector contains both a
string axion and a hidden sector axion, instead of an exactly massless
axion \textit{we actually expect a non-zero, suppressed mass for the
  axion}. Gravity is believed to not respect global symmetries (see,
e.g.,~\cite{Banks:2010zn}) and these violations may show up in a
low-energy effective theory as higher dimension operators that
explicitly break a global symmetry. Such explicit violations of the
hidden sector \(U(1)\) give a small, non-zero mass to the light
axion. An axion with a keV scale mass and \(f \approx 10^{15}\) GeV
together with a high supersymmetry breaking scale suggest an explicit
\(U(1)\) violating mass-squared operator suppressed by
\(1/M_{\text{pl}}^2f\) leading to an axion mass \(m \approx
\Lambda^4/M_{\text{pl}}^2f\).

The rest of this letter explores an explicit example of the general
scenario outlined above. We consider a string axion coupled to the
IYIT model of dynamical supersymmetry breaking~\cite{IY,*IT}. When the
scale of supersymmetry breaking is large, \(\Lambda_{\text{SUSY}}
\approx 10^{\rm 11-12}\) GeV, this model contains an axion that can
produce the 3.5 keV X-ray line seen in~\cite{exp1,exp2}.  As discussed
above, we believe the phenomena seen in our explicit example to be
common. For example, we note that it occurs in other
models of dynamical SUSY breaking such as the 4-1
model~\cite{4-1model}.

Following the demonstration of the 7 keV axion dark matter candidate,
we address potential cosmological issues that arise in our explicit
example. Some of these issues, such as isocurvature fluctuations, are
common to setups based on our general mechanism. However, we believe
the mechanisms employed to overcome certain cosmological issues in our
explicit example can be applied in more general scenarios.

We also include two appendices. In the first, we give in detail the
calculation of the axion spectrum for our explicit example. While the
techniques there can be found throughout the literature, we include
the derivation to keep our results self-contained. The second appendix
presents a new mechanism for dilaton stabilization. As a result of
this mechanism the axion develops an \(F\)-term; interestingly, this
effects the gaugino masses at the \(\mathcal{O}(1)\) level compared to
their values from anomaly mediation.

\section{An explicit model}

As an explicit realization of our setup, we consider the minimal IYIT
model~\cite{IY,*IT}. The model consists of four quark superfields
$Q^i$, \(i = 1, \dots, 4\), charged under a $Sp(1)\simeq SU(2)$ gauge
symmetry together with gauge singlets \(Z_{ij}\) in the \(\mathbf{6}\)
of the \(SU(4)\) flavor symmetry. Supersymmetric \(SU(2)\) gauge
dynamics lead to a quantum modified moduli space with
\(\mathrm{Pf}~(QQ) = \Lambda^4\), where \(\Lambda\) is the dynamical
scale of the theory. The gauge singlets couple to the quarks via a
tree-level superpotential \(W = \lambda Z_{ij} Q^iQ^j\). Supersymmetry
is broken by the $F$-term for $Z$, which cannot be simultaneously
satisfied with the quantum constraint.

The model contains a non-anomalous R-symmetry and an anomalous
\(U(1)_h\) symmetry under which \(Q\) and \(Z\) have charges \((0,1)\)
and \((2,-2)\), respectively. The \(U(1)_h\) symmetry, which has a
non-anomalous \(\mathbb{Z}_4\) subgroup, is spontaneously broken by
the quantum constraint. Therefore the phase of \(Q\) is the hidden
sector axion \(a_h\) with decay constant \(f_h \sim \Lambda\).

In addition to the fields \(Q\) and \(Z\), we consider a string axion
coupled to the gauge dynamics with strength \(1/(32\pi^2 f_s)\). The
\(SU(2)\) dynamical scale then contains the string axion
\(a_s\),
\begin{equation}
  \label{eq:iyitL}
  \Lambda = \mu e^{-\frac{8\pi^2}{b_0g^2}} e^{\frac{ia_s}{b_0f_s}} = |\Lambda| e^{\frac{ia_s}{b_0f_s}},
\end{equation}
where \(g\) is the gauge coupling and \(b_0\) is the coefficient of
the one-loop beta function. For \(Sp(N_c)\) gauge theories \(b_0 =
2(N_c+1)\), so for the scenario at hand \(b_0=4\). Presently, we
consider the dilaton and fermion partners of the string axion to be
stabilized and therefore non-dynamical. Otherwise, we would replace
\(ia_s\) by the chiral multiplet \(A_s\) in Eq.~\eqref{eq:iyitL}. In
Appendix~\ref{apx:dilaton} we present a possible mechanism of
stabilization.

The superpotential (with all indices suppressed) is
\begin{equation}
  \label{eq:iyitw}
  W = \lambda ZQQ + \frac{\mathcal{A}}{\Lambda^2}\left(\mathrm{Pf}~(QQ) - \Lambda^4\right),
\end{equation}
where the quantum constraint is enforced by the Lagrange multiplier
$\mathcal{A}$. In Appendix~\ref{apx:axion} we work out the effective
theory and axion spectrum in detail while keeping track of factors of
\(4\pi\) using na\"ive dimensional analysis
(NDA)~\cite{Luty:NDA1997,CKN:NDA1997}. However, it is simple to see
the basic results. Schematically, taking \(QQ \sim
\Lambda^2e^{2ia_h/f_h}\) and replacing \(\Lambda^4 \to
\Lambda^4e^{ia_s/f_s}\) in the quantum constraint, it is easy to see
that the \(F\)-term for \(\mathcal{A}\) produces a potential for the
axions \(a_h\) and \(a_s\),
\begin{equation}\label{eq:iyitpot}
V(a_h,a_s) \sim \Lambda^4 \left[1 - \cos\left(\frac{4a_h}{f_h} - \frac{a_s}{f_s} \right) \right].
\end{equation}
The above potential\footnote{The axion can be used as an inflaton in a
  chaotic inflation scenario since the K\"ahler potential for the
  axion has a shift symmetry~\cite{KYY}. In fact, the superpotential
  in Eq.~\eqref{eq:W_IYIT_eff_break} may generate a potential that is
  assumed in a chaotic natural inflation in supergravity proposed
  in~\cite{Linde}. The light axion has an effective decay constant
  \(b_0f_s\) so we need $b_0f_s >M_{\text{pl}}$ for successful natural
  inflation~\cite{Yonekura:2014,*Dine:2014}.} makes it clear that one
linear combination of axions gains a mass of order \(\Lambda^2/f_h\)
(for \(f_s\gg f_h\)) while the orthogonal combination is
massless.\footnote{Once a constant is added to the superpotential to
  cancel the cosmological constant, the field \(Z\) has a small
  expectation value \(\braket{Z} \sim m_{3/2}\). Therefore, the
  massless axion is, in fact, a linear combination of \(a_h, a_s,\)
  and the \(R\) axion (the phase of \(Z\)).}

As discussed previously, we generically expect quantum gravity to
violate the \(U(1)_h\) symmetry. Such explicit violations give the
massless axion from above a small, non-zero mass. To this end, we
consider the leading operator that violates the \(U(1)_h\) symmetry
while respecting the R-symmetry and the non-anomalous discrete
\(\mathbb{Z}_4 \subset U(1)_h\). With this criteria, the leading
operator is a deformation of the superpotential of the form\footnote{A
  K\"ahler operator of the form \(Z^*Z(QQ)^2/M_{\text{pl}}^4\) also
  respects the same symmetries and leads to the same order mass term
  as the operator in Eq.~\eqref{eq:iyitw2}.}
\begin{equation}
  \label{eq:iyitw2}
  W \supset \lambda'\frac{1}{M_{\text{pl}}^4}Z(QQ)^3 .
\end{equation}
We note that since the vacuum is located at \(\braket{Z} =
0\)~\cite{Chacko:IYIT1998} (see also Appendix~\ref{apx:axion}), the
above is the leading order term to the superpotential containing
\(Z\).

There are lower dimension operators that explicitly violate
\(U(1)_h\),
e.g.~\(\delta W = c \text{Pf}~(QQ) / M_{\text{pl}}\),
and therefore lead to different parametrics for the axion mass. These
operators violate the \(R\)-symmetry
and it is conceivable that this leads to their suppression,
e.g.~\(c \sim m_{3/2}/M_{\text{pl}} \sim \Lambda^2/M_{\text{pl}}^2\)
which gives a parametrically similar axion mass as the operator in
Eq.~\eqref{eq:iyitw2}. Thus we will consider only the operator of
Eq.~\eqref{eq:iyitw2} in the following analysis.

The explicit violation of the \(U(1)_h\) symmetry in
Eq.~\eqref{eq:iyitw2} gives a mass to the light axion through the
\(F\)-term for \(Z\) and is worked out in detail in the first
appendix. To leading order, the mass of the light axion is
\begin{equation}
  \label{eq:amass}
  m_a^2 \approx \frac{2 \lambda \lambda'}{(4\pi)^4} \frac{\Lambda^8}{M_{\text{pl}}^4f_s^2} = 2 \frac{\lambda\lambda'}{(\lambda/4\pi)^4}\frac{F^4}{M_{\text{pl}}^4f_s^2}.
\end{equation}
where \(F = \lambda \Lambda^2/(4\pi)^2\) is the scale of SUSY
breaking~\cite{Luty:NDA1997} (see also Appendix~\ref{apx:axion},
Eq.~\eqref{eq:F_SUSY}). As emphasized previously in a more general
context, here we explicitly see that the spectrum contains an axion
with a suppressed mass \(m_a \approx \Lambda^4/M_{\text{pl}}^2f_s\).

Through its string axion component, the light axion couples directly
to Standard Model photon operator $F\widetilde{F}$ with strength
\(1/f_s\). We can express the dynamical scale \(\Lambda\) in terms of
the decay rate,
\begin{equation}
  \label{eq:gammarate}
  \Gamma = \frac{\alpha_\mathrm{EM}^2}{64\pi^3}\frac{m_a^3}{f_s^2},
\end{equation}
as
\begin{equation}
  \label{eq:Lambdafinal}
\Lambda = \left( \frac{2\pi \alpha_{\text{EM}}^2 }{\lambda \lambda'} \frac{m_a^5M_{\text{pl}}^4}{\Gamma} \right)^{1/8}.
%  \sqrt{F} = \left( \frac{\alpha_{\text{EM}}^2}{2(4\pi)^7}\frac{\lambda^3}{\lambda'}\frac{m_a^5M_{\text{pl}}^4}{\Gamma} \right)^{1/8}.
\end{equation}
Experimental results~\cite{exp1, exp2} determine $m_a \approx 7$ keV and $\Gamma
\approx 5.7 \times 10^{-53}$ GeV. In the strongly-coupled vacuum the
coupling \(\lambda\) becomes non-perturbative and
\(\mathcal{O}(4\pi)\). Taking \(\lambda' \sim 1\), the supersymmetry
breaking scale is
\begin{equation}
  \label{eq:Lambdavalue}
  \sqrt{F} \sim 10^{11.5}~\mathrm{GeV},
\end{equation}
with a gravitino mass
\begin{equation}
  \label{eq:m3/2}
  m_{3/2} = \frac{F}{M_{\text{pl}}} \sim \mathcal{O}(10)\text{-}\mathcal{O}(100)~\mathrm{TeV}.
\end{equation}
We see that we have constructed an explicit model for the string axion
coupled to a hidden supersymmetry breaking sector where the scale of
supersymmetry breaking must be high to match the experimental X-ray
line.

We also know more about the spectrum of this model. The field $Z$ has
charge 2 under the $\mathbb{Z}_4$ symmetry; it cannot couple to
$W^\alpha W_\alpha$ to give the gauginos mass. Thus we are lead to
anomaly mediation (although see Appendix~\ref{apx:dilaton} for
modifications to the gaugino mass), which also fits nicely with the
gravitino mass above and the known Higgs mass. We can easily
incorporate this model in pure gravity
mediation~\cite{puregravity1,*puregravity2,*puregravity3} or minimal
split supersymmetry~\cite{splitsusy} models, to complete the extension
of the SM.

\section{Cosmology}

Although the recent observation of B-modes in the CMB by the BICEP2
collaboration~\cite{bicep2} has been shown to be consistent with 
expectations from dust~\cite{Flauger:2014qra,Adam:2014bub}, it has 
nevertheless renewed interest in models with a high inflationary scale, 
$H_{\text{inf}} \sim 10^{14}$ GeV, at the upper end of the currently-allowed 
range~\cite{Ade:2013uln}.  Large values of $H_{\text{inf}}$ present several 
cosmological challenges to any realistic model; for instance, isocurvature 
fluctuations of the nearly-massless axions must be suppressed. 
Furthermore, given a SUSY breaking scale of 
$\Lambda \sim 10^{11-12} \text{ GeV} < H_{\text{inf}}$, domain walls
are a potential problem due to the spontaneous breaking of the
$\mathbb{Z}_4$ symmetry after inflation.

It should be noted that the domain wall issue is a model-specific one,
which may be avoided by altering the dynamical SUSY breaking sector.
For example, the $\mathbb{Z}_4$ symmetry may be gauged, or a model
without a residual discrete symmetry may be chosen.  Of course, it may be 
that $H_{\text{inf}} < \Lambda$, although in this case isocurvature fluctuations may 
still pose a problem.  For the purposes of this section, we will focus on 
the model presented in the previous section and present a consistent 
cosmological history that addresses the aforementioned issues in the 
presence of a high inflationary scale.

For $\Lambda \sim 10^{12}$~GeV, as required by the analysis in the previous
section, the dynamical sector is weakly coupled during inflation. Thus
domain walls would be formed after reheating, once the temperature
fell below $\Lambda$. However, since $\Lambda$ is a dynamical scale,
we will show that it may be temporarily increased during
inflation. Taking $H_{\text{inf}} \lesssim \Lambda \lesssim
\rho_{\text{inf}}^{1/4}$ will ensure that the $\mathbb{Z}_4$ symmetry
is broken during inflation without the IYIT vacuum energy dominating
that of the inflaton.

Consider the gauge coupling of $SU(2)$ set by a gauge kinetic function
$f = \{(1/g_0^2) + c (\phi^2/M_{\text{pl}}^2)\}W^\alpha W_\alpha$ with
$W^\alpha$ the hidden sector gauge field strength, $\phi$ a singlet,
and $g_0$ a coupling set by string theory and compactification.  A
superpotential of the form $\kappa Y(\phi^2 - M_{\text{pl}}^2)$, for a
coupling $\kappa$ and a superfield $Y$, gives the singlet a large vev.
Generically, $\phi$ has a Hubble induced soft mass during inflation;
if $\kappa$ is sufficiently small, $\kappa \lesssim
H_{\text{inf}}/M_{\text{pl}}$, then $\braket{\phi} \sim 0$ and the
effective coupling $1/g^2 \approx 1/g_0^2$ is strong. The dynamical
scale, $\Lambda$, is easily of order $10^{15}$~GeV during inflation.
Domain walls are thus avoided as the $\mathbb{Z}_4$ symmetry is broken
during inflation by the IYIT meson condensate, so long as the
reheating temperature is sufficiently low, $T_R \lesssim \Lambda$.

Even with $\Lambda \sim 10^{15}$~GeV, the light axion is still essentially
massless compared to $H_{\text{inf}}$, leading to a potential
isocurvature problem.  This may be avoided by further increasing the
light axion mass during inflation, so that $m_a \sim
H_\text{inf}$.~\footnote{Note that the heavy axion mass, $m_{a'} \sim
  \Lambda$, is already heavy relative to $H_{\text{inf}}$ due to the
  impact of the $\phi$ singlet on the dynamical scale.}  This may be
accomplished by including a $U(1)_h$-violating coupling of the IYIT
quarks to the inflaton sector, giving an inflaton-field-dependent mass
to the light axion.  For concreteness, we assume a chaotic inflation
scenario as described in~\cite{KYY}.  Here the K\"ahler potential
respects a shift symmetry on the inflaton chiral multiplet, $H$, which
is broken by a mass term in the super potential, $W \supset m H X$.
We couple the IYIT quarks to the $X$ chiral multiplet in the K\"ahler
potential, $K \supset X^{\dag} X \text{Pf}\ (QQ) / M^4 + \text{h.c.}$
Once the dynamical SUSY breaking sector becomes strongly coupled, this
gives a mass term for $a_h$:
\begin{equation}
V \supset m^2 h^{\dag} h \left[1 - \frac{\Lambda^4}{M^4} \cos\left(\frac{4 a_h}{f_h} \right)\right].
\end{equation}
Here, $h$, the psuedo-scalar component of $H$, is the inflaton. Taking
$m^2 \left< h^{\dag} h \right> \sim H_{\text{inf}}^2 M_{\text{pl}}^2$
and $f_h \sim \Lambda$, we have $m_a \sim 4 H_{\text{inf}}
\left(\Lambda M_{P} / M^2\right)$.  Thus there will be no isocurvature
problem for $M \lesssim 10^{17}$ GeV.

Having given the light axion a large mass during inflation, it remains
to show that the appropriate misalignment can still be generated. If
the coupling, $c$, of the singlet to the gauge kinetic term is
complex, then the imaginary part may be removed by a shift in the
string axion, $a_s \rightarrow a_s + \text{Im}(c)
\braket{\phi}/M_{\text{pl}}$. When $\braket{\phi} = 0$, the axions
will find their minimum at the origin. After inflation, however, we
have $\braket{\phi} \sim M_{\text{pl}}$, and the heavy axion will
relax to its new minimum at $(4 a_h / f_h - a_s / f_s) \sim
\text{Im}(c)$. The effective potential for the light axion is then
approximately
\begin{equation}
V \sim \frac{2 \lambda \lambda'}{(4\pi)^4} \frac{\Lambda^8}{M_{\text{pl}}^4} \cos \left(2~\mathrm{Im}(c) + \frac{2 a_s}{f_s}\right).
\end{equation}
Since both axions were pinned to the origin during inflation, this
generates a misalignment of $a \sim \text{Im}(c) f_s$. In order that
this misalignment is sufficient to reproduce the observed abundance of
dark matter, we must have $\text{Im} (c) \sim 10^{-4}$.

Since $\phi$ was trapped at the origin during inflation, one may worry
that coherent oscillations of $\phi$ about its new minimum would come to dominate the energy density of the
Universe. Subsequent $\phi$ decays could lead to overproduction of
winos. However, if the Hubble induced mass remains significant as
$\phi$ relaxes, then $\phi$ will adiabatically roll to its minimum and
such oscillations do not occur. It is straightforward to check that,
at the time the inflaton decays, $\phi$ is displaced from its minimum
at $M_{\text{pl}}$ by an amount $\Delta \phi \sim H^2/(\kappa^2
M_{\text{pl}})$, where the Hubble constant \(H\) is evaluated at the
inflaton decay time. For a reheating temperature \(T_R\sim 10^9\) GeV
and \(\kappa \sim 10^{-5}\), \(\Delta \phi \sim 10^{-7}\) GeV. This
very small misalignment does not produce any appreciable amount of
$\phi$ oscillations.

One may further worry that domain walls are formed from the
spontaneous breaking of the \(\mathbb{Z}_2\) symmetry on \(\phi\) by
\(\braket{\phi}\ne 0\). These domain walls are an artifact of our
choice of the function of \(\phi\) in front of the gauge kinetic term
and in the superpotential. Other functions of \(\phi\) will do just as
well. In particular, we can add an explicit \(\mathbb{Z}_2\) breaking
term \(\epsilon \phi\) into the superpotential to collapse the domain
walls without changing our main results.

The dilaton superpartner to the string axion may similarly become
misaligned after inflation, depending on the exact form of its
K\"ahler potential. A mechanism such as that proposed
in~\cite{Linde_adiabatic,*Nakayama:2012mf} allows for the dilaton to
adiabatically roll to its minimum, and therefore its misalignment is
not dangerous.

\section{Discussion}

The 3.5 keV X-ray line observed in~\cite{exp1,exp2} allows the
exciting possibility that it may originate from dark matter. Tests of
this dark matter hypothesis are predominately limited to indirect
detection. Since the initial observation followup
studies~\cite{Malyshev:2014xqa, *Anderson:2014tza, *Tamura:2014mta}
have called into doubt this X-ray line. As there are now several
conflicting observations and the possible implications are far
reaching, it behooves us to continue to observe with future X-ray
experiments. As briefly studied in~\cite{exp1}, future X-ray
observations of galaxy clusters by the Astro-H Observatory can
distinguish the dark matter hypothesis from other astrophysical
sources that may be masquerading as a dark matter signal. We also note
that a promising place to look for a clean signal is from dwarf
spheroidal galaxies. These smaller galaxies can be dominated by
non-baryonic matter and yield a signal with smaller backgrounds. Even
if this 3.5 keV line is a red herring, the types of models we have
presented give an exciting link from a future X-ray signal to physics
at very high scales. This motivates further dedicated time to observe
these dwarf galaxies and galaxy clusters.

We briefly comment on other phenomenological consequences. For
example, the wino is stable and will generically be a sub-dominant
component of the dark matter in the Universe. It is in principle
possible to indirectly detect the wino dark matter utilizing gamma ray
observations of dwarf spheroidal galaxies or the galactic
center~\cite{wino_dwarf,*Cohen_wino,*Fan_wino}. However, it may be
very challenging since the wino indirect cross-section scales as the
density squared. As for the direct detection of the wino at the LHC,
see~\cite{atlas_wino}.

Embedding our mechanism into pure gravity mediation gives generic
predictions about the gluino mass. For example, for \(m_{3/2} \approx
50-100\) TeV then the gluino mass is in a detectable range at the
LHC. For a slightly larger gravitino mass, \(m_{3/2} \approx 200\)
TeV, we would expect \(m_{\tilde{g}} \approx 4-5\) TeV, and the gluino
would be hard to detect at the LHC. However, if the \(\mathcal{O}(1)\)
mass cancellation from the dilaton stabilization mechanism in
Appendix~\ref{apx:dilaton} takes place, then the gluino mass can be
\(m_{\tilde{g}} \approx 2-3\) TeV even for \(m_{3/2} \approx 200\)
TeV, and is therefore detectable.

Motivated by recent experimental results in the X-ray spectrum of
galaxy clusters and the current situation in particle physics beyond
the SM, we have explored the possibility of linking a keV signal to
supersymmetry breaking at a much higher scale, around $10^{11-12}$
GeV. This exciting experimental link between such different energy
scales is possible through a light axion, a mixture of a string theory
and a hidden (supersymmetry-breaking) sector axion, which gains only a
small mass from the supersymmetry breaking sector.

As an example demonstrating this possibility, we constructed a model
with dynamical supersymmetry breaking from the IYIT model coupled to a
string axion. One sees explicitly that there is a linear combination
of axions which does not gain a large mass, but only a small mass once
the superpotential includes higher dimensional operators. This mass is
directly related to the scale of the hidden sector, and thus
supersymmetry breaking. Using the X-ray results as input, we find the
scale of supersymmetry breaking to be $\mathcal{O}(10^{11.5})$
GeV. This scale fits nicely with models like pure gravity mediation or
minimal split supersymmetry.

Rather than producing a light axion to explain an X-ray signal, one
can instead construct similar models for the QCD axion. In this case
one needs to suppress operators to even higher dimension to produce a
lighter axion. Instead of using an $Sp(1)$ gauge group, a larger group
such as $Sp(5)$ should be used. Then we have a model for high-scale
supersymmetry breaking with a light QCD axion solving the strong CP
problem.

Given the lack of experimental evidence for supersymmetry thus far,
together with theoretical arguments for considering models which may
be difficult to see at colliders in the immediate future, it can be
fruitful to pursue new avenues for signals of supersymmetry. In this
letter we have shown a model which is well-motivated theoretically and
experimentally, and suggests a hint for supersymmetry in the keV sky.

\begin{acknowledgments}
  \noindent
  BH and DP would like to thank the hospitality of Kavli IPMU where
  this work was initiated. They are also grateful to Friends of
  UTokyo, Inc.~and WPI Initiative for travel support that helped make
  this work possible. The work of HM was supported by the U.S. DOE
  under Contract DE-AC03-76SF00098, by the NSF under grants
  PHY-1002399 and PHY-1316783, by the JSPS grant (C) 23540289. The
  work of TTY was supported by the Grant-in-Aid for Scientific
  Research on Innovative Areas, No.~26104009 and Scientific Research
  (B), No.~26287039. This work was supported by the World Premier
  International Research Center Initiative (WPI Initiative), MEXT,
  Japan.
\end{acknowledgments}

%\bibliography{kev_sa_refs}

\appendix
\onecolumngrid

% Appendix A, calculation details:
\input{./appendix.tex}

% Appendix B, dilaton
\input{./dilaton_stabilization.tex}

\twocolumngrid
\bibliography{kev_sa_refs}

\end{document}

%% file: appendix.tex
\section{Axion potential}\label{apx:axion}

In this appendix we derive the low-energy effective axion potential for the IYIT model discussed in the main text. In order to properly capture the axion dynamics---in particular, the mixing of the hidden sector axion and string axion---the Lagrange multiplier that enforces the \(SU(2)\) quantum constraint must be kept in the spectrum. This is because, in the absence of the string axion, the hidden sector axion and Lagrange multiplier pair up to become heavy together. Therefore, this analysis differs from the usual situation where the Lagrange multiplier is immediately integrated out of the spectrum, and we feel it is worthwhile, especially for non-experts of supersymmetric dynamics, to carefully lay out the steps of the calculation. In order to elucidate the physics, we first describe the simpler case of the model with no tree-level superpotential and no string axion and then add these terms to find the axion spectrum quoted in the text. 

\subsection{$W_{\text{tree}} = 0$}
We begin by considering \(SU(2)\) supersymmetric gauge theory with four quark superfields and no tree-level superpotential. Mesons \(M^{ij} = \epsilon^{\alpha\beta} Q_{\alpha}^iQ_{\beta}^j\)---in the \(\mathbf{6}\) of the \(SU(4)\) flavor symmetry---parameterize the moduli space of the low-energy supersymmetric vacua. In the quantum theory, instantons deform the moduli space and the mesons are subject to the constraint \(\text{Pf}~M = \Lambda^4\)~\cite{Seiberg:1994}, where \(\Lambda\) is the \(SU(2)\) dynamical scale and can always be made real by an anomalous \(U(1)_h\) rotation under which \(Q\) has unit charge. This constraint may be enforced in the low energy theory through a Lagrange multiplier, \(W = \mathcal{A}(\text{Pf}~M - \Lambda^4)\). In the following, we make use of the local Lie group isomorphism \(SU(4)\simeq SO(6)\) to describe the flavor symmetry; in this language, the mesons are in the vector representation of \(SO(6)\) and the quantum constraint is \(M_i^2 = \Lambda^4\).

Let us describe the qualitative features of the low-energy vacua. The quantum constraint spontaneously breaks the \(SO(6)\) flavor symmetry. At points of enhanced symmetry, the flavor symmetry is broken from \(SO(6) \to SO(5)\) by the vacuum expectation value
\begin{equation}
\braket{M_6} = \Lambda^2, \label{eq:vev}
\end{equation}
giving rise to five massless Nambu-Goldstone bosons. The quantum constraint also spontaneously breaks the anomalous \(U(1)_h\) symmetry under which \(Q\) has unit charge. The would-be NGB associated with the spontaneous breaking of \(U(1)_h\) gets a mass of order \(\Lambda\) through the anomaly and should be integrated out of the low energy theory. This would-be NGB, analogous to the \(\eta'\) meson of QCD, is what we refer to as the axion \(a_h\).

In summary, the quantum constraint, when satisfied as in Eq.~\eqref{eq:vev}, breaks the flavor symmetry as \(SO(6) \times U(1)_h \to SO(5)\) and five of the mesons are massless while the sixth one, the axion, gains a mass of order the dynamical scale. In the rest of this subsection, we show how this qualitative picture works out quantitatively in the effective theory. We then demonstrate how introducing a string-like axion leaves a massless axion in the low-energy theory.

In terms of the canonically normalized mesons \(\hat{M} \equiv
M/\Lambda\), the effective superpotential and K\"ahler potential are given by
\begin{subequations}
\label{eq:Leff_su2}
\begin{align}
W_{\text{eff}} &= \frac{1}{16\pi^2} \mathcal{A}\big( \hat{M}_i^2 - \Lambda^2\big) \label{eq:Weff_su2} \\
K_{\text{eff}} &= \frac{1}{16\pi^2}K_{\text{dyn}}(\hat{M}, \mathcal{A}) \label{eq:Keff_su2},
\end{align}
\end{subequations}
where \(\mathcal{A}\) is a Lagrange multiplier enforcing the quantum constraint. The factors of \(4\pi\) are included to ensure that the effective theory becomes strongly coupled at the scale \(\Lambda\) and are counted using na\"ive dimensional analysis~\cite{Luty:NDA1997, CKN:NDA1997}. We note that estimates using na\"ive dimensional analysis have an uncertainty factor of a few; we take this uncertainty to be implicit in our results and do not explicitly keep track of it. Since the quantum moduli space is smoothly described by the meson fields~\cite{Seiberg:1994}, we can take a canonical kinetic term for the meson fields as an apporiximation. Our results are not sensitive to this approximation. Further, a kinetic term for \(\mathcal{A}\) is generated at one-loop via the interaction \(\mathcal{A} \hat{M}^2\) in the effective superpotential. Therefore, at leading order the dynamical K\"ahler potential is given by
\begin{equation*}
K_{\text{dyn}}(\hat{M}, \mathcal{A}) \approx \hat{M}^{\dag}\hat{M} + \kappa \mathcal{A}^{\dag}\mathcal{A} %\label{eq:Kdyn_su2}
\end{equation*}
where \(\kappa \approx 5\) since there are five light mesons---the Nambu-Goldstone bosons---running in the loop that generates the kinetic term for \(\mathcal{A}\).

To study the \(SO(6)\times U(1)_h/SO(5)\) vacuum we parameterize the mesons as
\begin{equation} \label{eq:mes_param}
\hat{M} = e^{2 A_h/f_h}\Big(\hat{M}_a, \sqrt{\Lambda^2 - \hat{M}_a^2}\Big), \ a=1,\dots,5 .
\end{equation} 
The \(\hat{M}_a\) are the five NGB supermultiplets associated with the breaking \(SO(6) \to SO(5)\), while the axion supermultiplet \(A_h\), with scalar component \(s_h + ia_h\), is associated with the breaking of \(U(1)_h\). Inserting this parameterization into Eq.~\eqref{eq:Keff_su2}, expanding around small field values, and requiring a canonical kinetic term for the axion, we find the axion decay constant is given by \(f_h = \Lambda/(\sqrt{2}\pi)\). The superpotential is
\begin{equation}
W = \frac{\Lambda^2}{(4\pi)^2} \mathcal{A}\big( e^{4A_h/f_h} - 1 \big),
\end{equation}
and the \(F\)-term for \(\mathcal{A}\) gives the axion potential. In components, the vacuum lies at \(\braket{s_h} = 0\) and the axion potential is
\begin{equation}
V(a_h) = \frac{\Lambda^4}{8\pi^2\kappa}\left( 1 - \cos \frac{4a_h}{f_h} \right).
\end{equation}
The axion mass is easily seen to be
\begin{equation*}
m_{a_h}^2 = \frac{2\Lambda^4}{\kappa \pi^2 f_h^2} = \frac{4}{\kappa}\Lambda^2.
\end{equation*}

Now we consider an additional string axion coupled to the \(SU(2)\) gauge dynamics,
\begin{equation}
\mathcal{L} \supset - \int d^2\theta \frac{1}{16\pi^2} \frac{A_s}{f_s} \text{Tr}( W_{\alpha}^2) + \text{h.c.}, \ A_s = s_s + ia_s + \sqrt{2}\theta \psi_s + \dots 
\end{equation}
This coupling means that the \(SU(2)\) dynamical scale now depends on \(A_s\),
\begin{equation*}
\Lambda^4 \to \Lambda^4e^{A_s/f_s}.
\end{equation*}
It is a simple matter to find the effective potential including the string axion; we proceed exactly as above and find that the superpotential is
\begin{equation}\label{eq:W_As_Ah_mix}
W = \frac{\Lambda^2}{(4\pi)^2} \mathcal{A}\Big( e^{4A_h/f_h} - e^{A_s/f_s} \Big)
\end{equation}
Here we will assume the dilaton in \(A_s\) to be heavy and decoupled by some unspecified dynamics, while in Appendix~\ref{apx:dilaton} we present a new method to fix the dilaton. Then, the low energy axion potential is given by
\begin{equation}\label{eq:V_lagmult}
V(a_s,a_h) = \frac{\Lambda^4}{8\pi^2 \kappa} \left[ 1 - \cos \left( \frac{4a_h}{f_h} - \frac{a_s}{f_s} \right) \right].
\end{equation}
The above potential makes it clear that one linear combination of axions gets a mass of order \(\Lambda\) while the orthogonal direction is massless. It is a simple procedure to find the mass eigenstates; in the limit of \(f_s \gg f_h\) the heavy and light modes are given by:
%\begin{align}
%\text{Light } a: \quad & m_a^2 = 0, \quad a \approx a_s + \frac{f_h}{4f_s} a_h + \mathcal{O}\Big(\frac{f_h^2}{f_s^2}\Big), \quad f_a = \sqrt{16f_s^2 + f_h^2}\approx 4f_s  \\
%\text{Heavy } a': \quad & m_{a'}^2 = \frac{1}{8\pi^2 \kappa} \frac{\Lambda^4}{f_{a'}^2} \approx \frac{2\Lambda^4}{\kappa \pi^2 f_h^2}, \ a' \approx -a_h + \frac{f_h}{4f_s} a_s + \mathcal{O}\Big(\frac{f_h^2}{f_s^2}\Big), \ f_{a'} = \frac{\sqrt{16f_s^2 + f_h^2}}{f_sf_h} \approx \frac{4}{f_h} 
%\end{align}
\begin{equation}\label{eq:axion_masses1}
\begin{array}{cccc}
\text{Light } a: & m_a^2 = 0, & a \approx a_s + \dfrac{f_h}{4f_s} a_h + \mathcal{O}\Big(\dfrac{f_h^2}{f_s^2}\Big), & f_a = \sqrt{16f_s^2 + f_h^2}\approx 4f_s  \\
\text{Heavy } a': & m_{a'}^2 = \dfrac{1}{8\pi^2 \kappa} \dfrac{\Lambda^4}{f_{a'}^2} \approx \dfrac{2\Lambda^4}{\kappa \pi^2 f_h^2}, & a' \approx -a_h + \dfrac{f_h}{4f_s} a_s + \mathcal{O}\Big(\dfrac{f_h^2}{f_s^2}\Big), & f_{a'} = \dfrac{f_sf_h}{\sqrt{16f_s^2 + f_h^2}} \approx \dfrac{f_h}{4} 
\end{array}
\end{equation}
Note that the light axion picks up the larger decay constant, \(f_a \approx b_0 f_s\).

\subsection{$W_{\text{tree}} \ne 0$}
Now we consider the theory with the tree level superpotential considered in the text,
\begin{equation}
W_{\text{tree}} = W_{\text{IYIT}} + W_{\cancel{U(1)}},
\end{equation}
where \(W_{\text{IYIT}} = \lambda ZQQ\) spontaneously breaks supersymmetry and \(W_{\cancel{U(1)}}\) is a term which explicitly breaks the \(U(1)_h\) symmetry.

We first consider \(W_{\cancel{U(1)}} = 0\) and briefly review how the IYIT superpotential spontaneously breaks SUSY~\cite{IY,*IT} and the location of the vacuum~\cite{Chacko:IYIT1998}. For small \(Z\) the low-energy theory is still described by mesons and the effective superpotential is
\begin{equation}\label{eq:W_IYIT_eff}
W_{\text{eff}} = \frac{1}{(4\pi)^2} \Big[\lambda \Lambda Z_i \hat{M}_i + \mathcal{A}\big(\hat{M}_i^2 - \Lambda^2\big) \Big]
\end{equation}
where the singlets \(Z_i\) are in the \(\mathbf{6}\) of the \(SO(6)\) flavor symmetry. SUSY is broken through the \(F\)-term for \(Z\), which is incompatible with the quantum constraint \(\hat{M}_i^2 = \Lambda^2\). The vacuum lies in the direction where the \(SO(6)\) symmetry is broken to \(SO(5)\); this gives rise to five massless NGBs while their supersymmetric scalar partners gain a SUSY breaking mass of order \(\lambda \Lambda\). Note that the singlets have canonical kinetic terms, \(K_{\text{eff}} \supset Z^{\dag}Z\) (with no factors of \(4\pi\)). Therefore the SUSY breaking scale is suppressed from the dynamical scale \(\Lambda\) by extra factors of \(4\pi\)~\cite{Luty:NDA1997}
\begin{equation}\label{eq:F_SUSY}
F = \frac{\lambda}{(4\pi)^2} \Lambda^2 .
\end{equation}

The superpotential in Eq.~\eqref{eq:W_IYIT_eff} is an O'Raifeartaigh model of SUSY breaking and therefore comes with a classically flat direction; namely, in the vacuum \(\braket{\hat{M}_6} = \Lambda\) the singlet \(Z_6\) is massless at tree-level and its value is undetermined. For perturbative values of the coupling \(\lambda\), the theory is calculable near the origin and one finds that there is a stable, local minimum located at \(\braket{Z_6} = 0\)~\cite{Chacko:IYIT1998}.\footnote{If we explicitly keep the heavy axion and Lagrange multiplier \(\mathcal{A}\) in the effective theory, as in the meson parameterization in Eq.~\eqref{eq:mes_param}, then the classically flat direction is not \(Z_6\) but instead it is a linear combination of \(Z_6\) and \(\mathcal{A}\). The results of the preceding paragraph and reference~\cite{Chacko:IYIT1998} remain the same for this classically flat direction.}

Turning on the explicit \(U(1)_h\) breaking term, \(W_{\cancel{U(1)}}\), the light axion in Eq.~\eqref{eq:axion_masses1} will gain a small, non-zero mass. It is simple to estimate the size of this mass; the explicit breaking term is of the form
\[
W_{\cancel{U(1)}} = \frac{\lambda'}{M_{\text{pl}}^n} \mathcal{O}_{\cancel{U(1)}}
\]
with \(\mathcal{O}_{\cancel{U(1)}}\) a dimension \(n+3\) operator that explicitly breaks \(U(1)_h\). The light axion carries the string decay constant \(f_s\) (see Eq.~\eqref{eq:axion_masses1}) and the only other dimensionful scales in the problem are \(M_{\text{pl}}\) and \(\Lambda\) (since we are in the vacuum where \(\braket{Z_6} = 0\)). Therefore, up to numerical factors, the light axion mass is
\[
m_a^2 \sim \lambda'\frac{\Lambda^{n+4}}{M_{\text{pl}}^nf_s^2}.
\]
In the text we consider the leading operator that breaks \(U(1)_h\) while preserving \(U(1)_R\), \(\mathcal{O}_{\cancel{U(1)}} \sim \lambda'Z(QQ)^3 \sim \Lambda^3\lambda'^{ijkl}Z_i\hat{M}_j\hat{M}_k\hat{M}_l\). For simplicity, we may take the \(SO(6)\) preserving interaction with \(\lambda'^{ijkl} = \lambda' \delta^{ij} \delta^{kl}\)---flavor violating effects can be considered as perturbations around this and do not significantly change the results. Thus, we examine the effective superpotential
\begin{equation}
\label{eq:W_IYIT_eff_break}
W_{\text{eff}} = \frac{1}{(4\pi)^2} \Big[\lambda \Lambda Z_i \hat{M}_i + \lambda'\frac{\Lambda^3}{M_{\text{pl}}^4}Z_i\hat{M}_i\hat{M_j}^2 + \mathcal{A}\big(\hat{M}_i^2 - \Lambda^2e^{A_s/f_s}\big) \Big].
\end{equation}
It is straightforward to compute the axion potential; there is a contribution from \(\mathcal{A}\)'s \(F\)-term (Eq.~\eqref{eq:V_lagmult}) while the explicit breaking manifests itself in the \(F\)-term for \(Z\) giving
\begin{equation}\label{eq:V_explicit}
\delta V(a_h,a_s) = \frac{\Lambda^4}{(4\pi)^4}\left[\lambda^2 + \lambda'^2\frac{\Lambda^8}{M_{\text{pl}}^8} +2\lambda \lambda' \frac{\Lambda^4}{M_{\text{pl}}^4} \cos\left(\frac{4a_h}{f_h}\right) \right]
\end{equation}
The full axion potential is given by Eq.~\eqref{eq:V_lagmult} plus the above contribution. To leading order in \(f_h/f_s\) the mass of the heavy axion is unchanged from Eq.~\eqref{eq:axion_masses1} while the light axion gains a mass of
\begin{equation}
m_a^2 \approx \frac{2 \lambda \lambda'}{(4\pi)^4} \frac{\Lambda^8}{M_{\text{pl}}^4f_s^2} = \frac{2\lambda \lambda'}{(\lambda/4\pi)^4}\frac{F^4}{M_{\text{pl}}^4f_s^2},
\end{equation}
where in the last equality we expressed the mass in terms of the SUSY breaking scale in Eq.~\eqref{eq:F_SUSY}.

Finally, we comment on the value of the yukawa coupling \(\lambda\). In the strongly coupled vacuum, \(\lambda\) quickly becomes non-perturbative as the wave-function for the quarks shrinks to zero. For \(\lambda = 4\pi\) we cannot prove that \(\braket{Z} = 0\) is a stable minimum~\cite{Chacko:IYIT1998}. Thus we must assume it is the case.

%% file: dilaton_stabilization.tex
\section{Dilaton Stabilization}\label{apx:dilaton}

Dilaton stabilization, or more generally, moduli stabilization, is a notorious problem in string theory. When the string axion couples to the sector responsible for dynamical SUSY breaking there are several typical issues. First, there is a runaway direction in which SUSY is preserved. More concretely, the vacuum energy is order \(V \sim \Lambda^4\) where \(\Lambda\) is the dynamical scale of the hidden sector. With the string axion, this is modified to \(V \sim \Lambda^4 e^{(\phi+\phi^*)/f_s}\), where \(\phi\) is the scalar modulus of the string axion supermultiplet so that \(s \equiv \phi + \phi^*\) is the dilaton. Clearly, the potential is minimized for \(s \to-\infty\) with \(V \to 0\) and SUSY is restored.

There are ways to stabilize moduli, such as KKLT~\cite{kklt} or racetrack scenarios~\cite{Krasnikov:1987jj,*Dixon:1990ds,*Casas:1990qi}. However, these are supersymmetric preserving mechanisms, so both the dilaton and the string axion are fixed. One may want, as in this work, a mechanism which stabilizes the dilaton but leaves the string axion free. This is clearly a non-supersymmetric request, and therefore any such mechanism that achieves this must make use of the dynamical supersymmetry breaking sector or some other source of supersymmetry breaking. In these setups, the dilaton typically has a gravitino sized mass, \(m_{3/2}\). Such scenarios are not easily constructed, and they typically have other issues, such as so-called dilaton domination of SUSY breaking~\cite{dilaton_dom}. Here, due to the dilaton's coupling to the Standard Model gauge sector, gauginos get very large masses, \(m_{1/2} \gtrsim m_{3/2}\). In anomaly mediated SUSY breaking \(m_{3/2} \sim \mathcal{O}(100)\) TeV, making the phenomenology uninteresting.

In this appendix, we suggest a novel mechanism that both stabilizes the dilaton and does not introduce a large gaugino mass despite a large $F$ term for the dilaton. As we will see, the gaugino mass coming from the dilaton is comparable to its mass from anomaly mediation. This means that the anomaly mediated mass may be \(\mathcal{O}(1)\) changed, which is exciting in its own right. The key observation in this setup is to take both the modulus and the IYIT singlet fields \(Z\) to live in a strongly coupled sector~\cite{strong_moduli}. By analogy with composite models, we make the crucial assumption that a form of na\"ive dimensional analysis (NDA) also holds for these fields. Properly counting the factors of \(4\pi\) coming from NDA then give the results outlined above.

A comment on notation: in this appendix \(\phi\) refers to a modulus field like the string axion supermultiplet. In relation to string axion multiplet \(A_s\) and the hidden sector multiplet \(A_h\) considered in the text, \(\phi\) is the linear combination of them that is light.\footnote{In Eq.~\eqref{eq:W_As_Ah_mix} and the following discussion it is easy to see that an entire chiral superfield is left massless if the dilaton is not fixed. Note that this linear combination generically picks up the larger decay constant (\textit{e.g.} section VI.F.4 of~\cite{Kim_2008_axion_review}), \textit{i.e.} the string decay constant (see also Eq.~\eqref{eq:axion_masses1}). Therefore this \(\phi\) really does behave like a string modulus and the results of this section apply more generally.} For clarity, we ignore \(\mathcal{O}(1)\) constants such as the one-loop beta function coefficient \(b_0\); these are easily restored and do not alter our results.
 
We assume that there is a K\"ahler coupling between the string axion multiplet and the singlet fields,
\begin{subequations}
\begin{align}
K &\supset g(\phi+ \phi^*) + h(\phi + \phi^*)Z^{\dag}Z \\
&\approx M(\phi + \phi^*) + \frac{1}{2}(\phi+\phi^*)^2 + \dots + Z^{\dag}Z\left( 1 + \frac{\phi+\phi^*}{M} + \dots \right),
\end{align}
\end{subequations}
%where \(M = 32\pi^2 f_s \sim 10^{18}\) GeV for \(f_s \sim 10^{15}\) GeV.
where the scale \(M\) corresponds to the Plank scale, \(M \simeq 2.4 \times 10^{18}\) GeV. In this case, the vacuum energy is given by
\begin{equation}
\label{eq:V_dil}
V = K^{Z Z^{\dag}}\left|\frac{\partial W}{\partial Z} \right|^2 \sim \frac{\lambda^2 \Lambda^4 e^{s/f_s}}{1 + \frac{s}{M} + \dots}.
\end{equation}
The potential is minimized for \(s = -M + f_s \sim -M\) with vacuum energy \(V \sim \lambda^2 \Lambda^4 e^{-M/f_s}/(f_s/M)\). While the dilaton is technically stabilized and SUSY is still broken, the vacuum energy is tiny for the typical string axion parameters we are considering as it is suppressed by \(e^{-M/f_s} \sim e^{-10^3}\).

Moreover, the dilaton acquires a large \(F\)-term, \(F_{\phi} \sim M m_{3/2}\):
\begin{equation}
F_{\phi} \approx K^{\phi\phi^*}\Big(W_{\phi} + K_{\phi} W/M_{\text{pl}}^2\Big) \approx \frac{K_{\phi} W}{M_{\text{pl}}^2} \approx M m_{3/2}.
\end{equation}
where we have taken the superpotential to contain a constant so as to cancel the cosmological constant, \(W \sim Z\Lambda^2 e^{\phi/f_s} + m_{3/2}M_{\text{pl}}^2\), and evaluated \(F_{\phi}\) in the vacuum \(\braket{Z} = 0\). As the string axion couples to the Standard Model sector, this \(F\)-term leads to a large gaugino mass,
\begin{equation}
\label{eq:gaugino_mass1}
\int d^2\theta \frac{\phi}{32\pi^2f_s} W_{\text{SM},\alpha}W_{\text{SM}}^{\alpha} \Rightarrow m_{1/2} \sim \frac{M}{32\pi^2 f_s} m_{3/2} \gg m_{3/2},
\end{equation}
with \(32\pi^2f_s \sim 10^{17}\) GeV for \(f_s \sim 10^{15}\) GeV.

These results change if dilaton sector is strongly coupled. We then imagine that the proper low-energy effective theory becomes strongly coupled at the compactification scale. We are led to consider the string scale as a composite scale, which we label \(M_c\), and apply the rules of na\"ive dimensional analysis.

%Luty:NDA1997,CKN:NDA1997,
Let us briefly review the rules of NDA~\cite{Georgi_NDA1,*Georgi_NDA2,Luty:NDA1997,CKN:NDA1997}: multiply the effective action by an overall factor of \(1/16\pi^2\), replace the composite fields as \(\Phi \to 4\pi \Phi\), and relabel the cutoff \(M \to M_c\). For example, the coupling of the dilaton to SM gauge fields in Eq.~\eqref{eq:gaugino_mass1} becomes
\begin{equation}
\label{eq:gauge_kin_NDA}
\frac{\phi}{M} W_{\text{SM},\alpha}W_{\text{SM}}^{\alpha} \to \frac{4\pi \phi}{16\pi^2 M_c} W_{\text{SM},\alpha}W_{\text{SM}}^{\alpha}.
\end{equation}
As the above operator is responsible for the string axion decay to photons considered in the text, it sets the scale \(M_c\):
\begin{equation}
\int d^2\theta \frac{\phi}{4\pi M_c} W_{\text{EM},\alpha}W_{\text{EM}}^{\alpha} \supset \frac{a}{32\pi^2 f_s}F_{\text{EM}}\tilde{F}_{\text{EM}},
\end{equation}
so that for the observed value \(f_s \sim 10^{15}\) GeV we have \(M_c \sim 10^{16}\) GeV.

Let us assume that the singlet fields \(Z\) also live in the strongly coupled sector; then we can view them as ``composite'' particles just like the string axion. Using NDA, the relevant terms in the effective K\"ahler potential are
\begin{equation}
K \sim \frac{1}{16\pi^2} \left[ 4\pi M_c (\phi + \phi^*) + \frac{(4\pi)^3}{M_c} Z^{\dag}Z(\phi + \phi^*) \right].
\end{equation}
The vacuum energy as a function of the dilaton is then of the same form as Eq.~\eqref{eq:V_dil} with \(M = M_c/4\pi\) and \(f_s = M_c/8\pi\). Since \(M \sim f_s \sim 0.1 M_c\), we immediately see that the vacuum energy is no longer tiny: at the minimum \(V'(s) = 0\) we have \(s = - M + f_s \sim -f_s \sim -M_c/8\pi\) so that the vacuum energy is \(V \sim \lambda^2 \Lambda^4 e^{-1}/\mathcal{O}(1)\). The dilaton F-term is \(F_{\phi} = M_c m_{3/2}/4\pi \approx f_s m_{3/2}\); this gives the gauginos a mass of order
\begin{equation}
\label{eq:gaugino_mass2}
m_{1/2} \approx \frac{F_{\phi}}{32\pi^2f_s}m_{3/2} \approx \frac{1}{32\pi^2} m_{3/2}.
\end{equation}
This contribution to the gaugino mass from the dilaton is comparable in size to the gaugino mass coming from anomaly mediation.

In summary, we have outlined a mechanism to stabilize the dilaton---while leaving the string axion free---that is phenomenoligically viable with supersymmetry breaking and a string axion decay constant that could explain the 3.5 keV line, as described in the main text. Moreover, the stabilization mechanism may allow for the anomaly mediation relations for gaugino masses to be changed by an \(\mathcal{O}(1)\) amount, which could prove useful for model building. The crucial assumption in achieving these results is that the effective action of the dilaton should become strongly coupled at the compactification scale. This led us to applying na\"ive dimensional analysis to ensure this behavior of the effective action. By also considering the singlets involved in SUSY breaking to live in the strongly coupled sector we obtain the stated results.